\documentclass[12pt]{article}
\usepackage{graphicx}
\usepackage{color}
\definecolor{darkred}{rgb}{0.75,0.0,0.0}
\newcommand\beq{\begin{equation}}
\newcommand\eeq{\end{equation}}
\newcommand\bear{\begin{eqnarray}}
\newcommand\eear{\end{eqnarray}}
\begin{document}
\baselineskip=24pt

\begin{center}
{\Large \bf Half-Metallicity in Undoped and Boron Doped Graphene Nanoribbons
in Presence of Semi-local Exchange-Correlation Interactions} 
\end{center}

\vspace*{0.5cm}

\centerline{\bf Sudipta Dutta and Swapan K. Pati$^*$}

\vspace*{0.1cm}

\begin{center}
{Theoretical Sciences Unit and DST unit on Nanoscience
\\Jawaharlal Nehru Centre For Advanced Scientific Research
\\Jakkur Campus, Bangalore 560064, India.}
\end{center}

\begin{center}
{\bf Abstract}
\end{center}

\vspace*{0.2cm}

{We perform density functional calculations on one-dimensional 
zigzag edge graphene nano-ribbons (ZGNRs) of different widths, with and 
without edge doping including semi-local exchange-correlations. Our study 
reveals that, although the ground state of edge passivated (with hydrogen) 
ZGNRs prefers to be anti-ferromagnetic, the doping of both the edges with 
Boron atoms stabilizes the system in a ferromagnetic ground state. Both 
the local and semi-local exchange-correlations result in half-metallicity 
in edge passivated ZGNRs at a finite cross-ribbon electric field. 
However, the ZGNR with Boron edges shows
half-metallic behavior irrespective of the ribbon-width even in absence 
of electric field and this property sustains for any field strength, 
opening a huge possibility of applications in spintronics.}

\newpage
\clearpage

Nanomaterials of Carbon of different dimensionalities like Carbon nanotubes, 
fullerenes etc. have been a subject of interest over the past few decades 
due to their potential applications in various nanoscale electronic 
devices\cite{Dresselhaus,Chico,McEuen}. Graphene, the two dimensional flat 
monolayer of Carbon atoms packed into a honeycomb lattice is the latest 
addition to this family owing to the recent progress in experimental 
techniques\cite{Novo1,Novo2,Zhang,Stankovich}. Because of its sophisticated 
low-dimensional electronic properties and huge application possibility, it 
has attracted the focus of a big scientific community in recent times to 
explore it in various 
aspects\cite{Meyer,Novo3,Novo4,Katsnelson,Schniepp,Jiang,Agapito,Grimme}. 
   
The electronic properties of nanoscale Carbon systems are governed by their 
size and geometry. Recent progress in experiments allows to make finite size 
graphene layer, termed as graphene nanoribbons (GNRs) with varying widths, 
either by cutting mechanically exfoliated graphenes\cite{Novo1,Novo2,Zhang} 
and patterning by electron beam lithography\cite{Ozyilmaz}, or by controlling 
the epitaxial growth of graphenes\cite{Berger1,Berger2}. Different 
possibilities of geometrical termination of graphene layer give rise to two 
different edge geometries, namely, zigzag and armchair edges, differing 
largely in their electronic properties. These different edge geometries have 
been modeled by imposing different boundary conditions on Schrodinger's 
equation within the tight-binding limit\cite{Fujita,Nakada,Wakabayashi,Ezawa} 
or on the Dirac equation for two-dimensional massless fermions with an effective 
speed of light\cite{Brey,Sasaki,Abanin,Neto} in previous studies. There exists 
a few many body studies also on the low-dimensional electronic 
properties\cite{Sudipta,McCann,Yang,Prezzi,Lin} of these systems. These also 
have been extensively studied using density functional 
theory\cite{Cohen,Okada}.

One-dimensional antiferromagnetic zigzag edge graphene nano-ribbons (ZGNRs) 
show half-metallicity at a finite external electric field across the ribbon 
width within local density approximation (LDA) as observed by Son 
{\it et. al.}\cite{Cohen}. Half-metallic materials show zero band gap for 
electrons with one spin orientation and semiconducting or insulating band gap 
for the other, giving rise to completely spin polarized current, which has 
already been observed in some materials like Heusler compounds\cite{Groot}, 
manganese perovskite\cite{Park} etc. However, one recent study predicted that, 
in presence of non-local exchange-correlations like B3LYP, the cross-ribbon 
electric field makes the band gap of one spin orientation insulating and the 
other semiconducting for a finite size edge passivated ZGNR with possible 
application as spin-selective semiconductors\cite{Rudberg}.

To enlighten the controversy on the half-metallicity of ZGNRs in presence 
of non-local exchange-correlations, we study 1D periodic ZGNRs of different 
widths with a variety of edge passivation and doping using {\it ab initio} density 
functional package, SIESTA\cite{Soler}. We have performed spin polarized 
calculations within generalized gradient approximation (GGA) considering 
Perdew-Burke-Ernzerhof (PBE) exchange and correlation functionals\cite{Perdew} 
with double zeta polarized (DZP) basis set. GGA approximation takes into 
account the semi-local exchange-correlations which have significant impact on 
low-dimensional spin systems like GNRs. The 1D ZGNRs along $x$-axis and in 
$xy$-plane are represented by the unit cell lattice vectors $2.45, 30.0, 16.0$, 
which create sufficient vacuum, ensuring no interaction between the adjacent 
layers\cite{Cohen}. We have considered $400$ Rydbergs energy cut-off for a real 
space mesh size and a k-point sampling of 36 k-points, uniformly positioned 
along the 1D Brillouin zone. The systems considered are depicted in Fig.1 with 
$(i)$ all the edge Carbon atoms passivated by hydrogen and $(ii)$ all the edge 
Carbons replaced (doped) with Boron (B). We vary the ribbon width ($w$) for both 
the cases from $4$ to $10$ zigzag chains, and relax all the structures for 
both ferromagnetic and antiferoomagnetic spin orientations. We then apply a
transverse electric field along the cross-ribbon direction, i.e., along the 
$y$-axis as shown in Fig.1 and track the band gap 
variation for both the spins orientations.

\begin{figure}
\centering
\includegraphics[scale=0.5, angle=0] {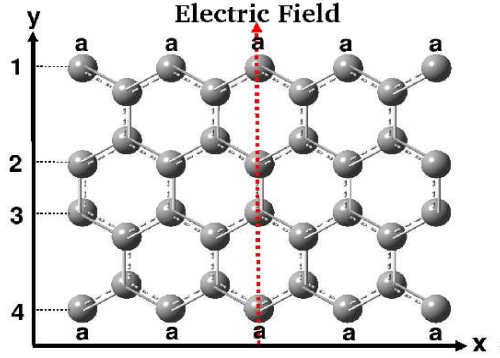}
\caption{(Color online) Model system: a zigzag edge 1D graphene with 4 zigzag
chains (4-ZGNR), translated along $x$-axis. The edge atoms are denoted by "a".
The dotted arrow shows the direction of external electric field.}
\end{figure}

\begin{figure}
\centering
\includegraphics[scale=0.8, angle=0] {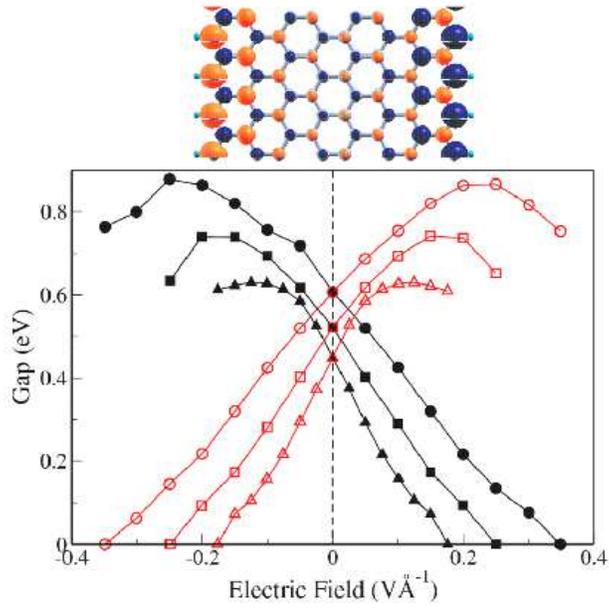}
\caption{(Color online) Top picture shows the accumulation of up spin and down spin
densities on either edges of a finite segment of 1D $8$-ZGNR with hydrogen passivated
edges in absence of electric field. The down panel shows the up spin (closed symbols)
and down spin (open symbols) gaps for $6$ (circle), $8$ (square) and $10$-ZGNRs
(triangle) as a function of electric field. Negative field corresponds to reverse 
direction.}
\end{figure}

The undoped ZGNRs with hydrogen-passivated edges show that the antiferromagnetic 
ground state is stabler than the ferromagnetic state by a small margin, of the 
order of meV. The ground states with zero magnetic moment show antiparallel 
alignment of spins at opposite edges as well as between nearest neighbors, as 
depicted in Fig.2. These observations are consistent with the previous LDA 
results\cite{Cohen} as well as with the theory based on the Hubbard Hamiltonian 
for bipartite lattice\cite{Lieb}. However, the electron-electron interactions 
can overcome such a small gap to make the ground state of ZGNRs 
ferromagnetic\cite{Sudipta}. In absence of external electric field, both the 
upspin and downspin band gaps show insulating behavior and the insulating gap 
decreases with the increase in width, as can be seen from Fig.2. The gap appears
to be more than the previously reported LDA gap, since LDA is known to 
underestimate the band gap. The external electric field perturbs the ground 
state band structure. Increase in field strength along 
positive direction decreases the band gap for the upspin and at a finite electric 
field strength, the up-spin channel becomes metallic. The band gap for down-spin 
however, increases with the field strength. As a result, the 1D ZGNRs become 
half-metallic 
beyond a finite critical field strength, termed as $E_{c,w}$. From Fig.2 it is 
clear that, the $E_{c,w}$ decreases with the increase in width ($w$). Electric 
field in reverse direction closes the down-spin gap and increases the up-spin 
insulating gap. Thus, different directions of external electric field open up 
different spin channels. Since, the edges have different spin polarizations, the 
applied bias $eEy$ ($e$, $E$ and $y$ are electronic charge, electric field and 
the distance respectively) in either directions affects the up-spin and down-spin 
band gaps in opposite fashions. Our results thus suggest that in presence of 
gradient corrected semi-local exchange-correlations, 
the 1D ZGNRs do behave as half-metallic materials beyond a certain electric field 
strength. This result strongly contradicts the observation of spin-selected 
semiconducting behavior of ZGNRs in presence of non-local exchange- 
correlation as found by Rudberg et al\cite{Rudberg}. The controversy of semiconducting gap for one spin 
channel arises due to the finite size of the system, used in ref\cite{Rudberg} and such 
a gap can be tuned with increase in length (along $x$-axis) of ZGNRs. 
The gradual spin density annihilation at the edges of finite size ZGNRs with
increase in electric field as observed by Rudberg et al\cite{Rudberg} in presence of non-local 
exchange-correlation is nevertheless due to finite size effect\cite{Kan}. 
Moreover, very high field strength 
as considered in ref\cite{Rudberg} is unphysical (beyond perturbing 
regime) and leads to collapse of the structure of such a nano-scale system.

The alternate idea of edge passivation is to dope the edge atoms by three 
coordinated Boron atoms, with introduction of holes in the system. We find that
B doping at the edges makes the ground state of ZGNRs ferromagnetic. Such a 
possibility with large hole doping in strongly correlated low-dimensional systems 
was proposed earlier\cite{HRK}. In our case, the energy difference between the 
ferromagnetic and antiferromagnetic ground state is of the order of meV. In the 
ground state, the electrons at the same edges as well as at the opposite edges 
are parallely aligned (see Fig.3). In Fig.4, we plot the density of states (DOS) 
for this system for zero and nonzero electric field. As can be seen, even in 
absence of any external electric field, this system shows metallic behavior 
for the majority spin and insulating behavior for the minority spin, the typical 
half-metallic behavior. The increase in field strength in either directions does 
not change this behavior: the majority spin channel always remains conducting and 
the minority spin gap remains insulating with almost same gap for all the field 
strengths (see Fig.4). It is because, the external field in either directions 
affects both the spins channels similarly, since the spin polarization of the 
opposite edges are same in this system. As a consequence, the change in 
direction of the external field cannot change the spin polarization of the current.
The minority spin gaps for all the widths are much higher compared to the room 
temperature ($0.025$ eV). Thus, the ZGNRs with both the edges doped with B can act 
as a potential half-metallic material for any external electric field over a large 
temperature domain irrespective of its width. The projected DOS (pDOS) analysis shows 
that, the half metallicity at Fermi energy in this system is largely induced by the 
B orbitals at the edges.  

\begin{figure}
\centering
\includegraphics[scale=0.15, angle=0] {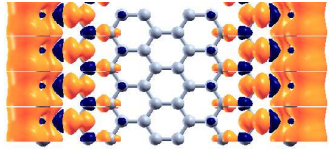}
\caption{(Color online) Accumulation of up spin densities on both the edges of a
finite segment of 1D $8$-ZGNR with all the edge atoms doped with Boron in absence
of electric field.}
\end{figure}

\begin{figure}
\centering
\includegraphics[scale=0.3, angle=270] {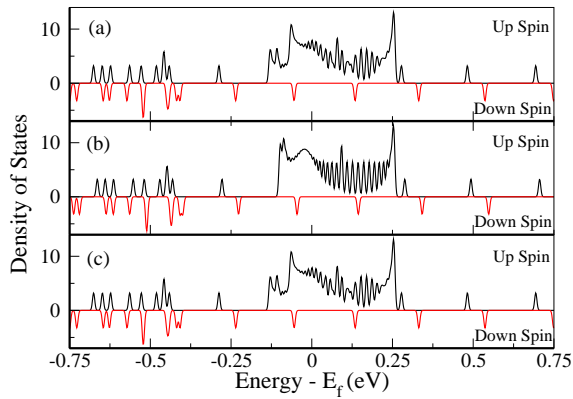}
\caption{(Color online) Up and down spin density of states of $8$-ZGNR with both
the edges doped by Boron as a function of energy, scaled with respect to the
Fermi energy for three different electric fields $(a) -0.2$, $(b) 0.0$ and
$(c)0.2$ $V \AA^{-1}$.}
\end{figure}

We have also studied the ZGNRs of several widths with both the edges doped by 
Nitrogen (N) atoms. This system shows antiferromagnetic ground state with 
different spin polarization of the opposite edges. However, although the electron 
accumulation on the edges is noticably low compared to the systems discussed above, 
this system shows metallic behavior for both the spins at Fermi energy even in 
absence of external electric field. The pDOS analysis shows that, the N orbitals 
contribute largely to the DOS at Fermi energy making the system metallic. All 
these observations remain consistent with the increase in width.

In summary, we have studied the 1D ZGNRs of different widths with edge passivation 
and with edge doping in presence of transverse electric field. Our study shows that, 
even in presence of gradient corrected semi-local exchange-correlations, the edge 
passivated 1D ZGNRs 
can act as half-metallic materials beyond a certain critical field strength, which 
decreases with the increase in ribbon width. However, Boron doped ZGNRs show 
half-metallic behavior even in absence of external electric field and this 
half-metallicity sustains for any finite field strength over a large temperature 
range. This observation opens a huge possibility of application of B doped ZGNRs in 
spintronic devices.

Acknowledgement: SD acknowledges the CSIR for the research fellowship and SKP acknowledges 
the research support from DST and CSIR, Govt. of India.


\begin{thebibliography}{50}

\bibitem{Dresselhaus} Dresselhaus, G.; Dresselhaus M. S.; Eklund, P. C.
Science of Fullerenes and Carbon Nanotubes: Their Properties and Applications
(Academic, New York,  {\bf 1996}).
\bibitem{Chico} Chico, L.; Crespi, V. H.; Benedict, L. X.; Louie, S. G.;
Cohen, M. L. {\it Phys. Rev. Lett.} {\bf 1996}, {\it 76},  971.
\bibitem{McEuen} McEuen, P. L.; Fuhrer, M. S.; Park, H. {\it IEEE Trans. Nanotechnol.} 
{\bf 2002}, {\it 1}, 78.
\bibitem{Novo1} Novoselov, K. S. {\it et. al.} {\it Proc. Natl. Acad. Sci. U. S. A.}  
{\bf 2005}, {\it 102}, 10451.
\bibitem{Novo2} Novoselov, K. S. {\it et. al.} {\it Nature (London)} {\bf 2005}, {\it 438},  
197.
\bibitem{Zhang} Zhang, Y.; Tan, Y. W.;  Stormer, H. L.; Kim, P. {\it Nature (London)} 
{\bf 2005}, {\it 438}, 201.
\bibitem{Stankovich} Stankovich, S. {\it et. al.} {\it Nature (London)} {\bf 2006}, {\it 442},  
282.
\bibitem{Meyer} Meyer, J. C. {\it et. al.} {\it Nature} {\bf 2007}, {\it 446}, 60.
\bibitem{Novo3} Novoselov, K. S. {\it et. al.} {\it Science} {\bf 2007}, {\it 315}, 1379.
\bibitem{Novo4} Geim, A. K.; Novoselov, K. S. {\it Nature Materials} {\bf 2007}, {\it 6}, 183.
\bibitem{Katsnelson} Katsnelson, M. I. {\it Mater Today} {\bf 2007}, {\it 10}, 20.
\bibitem{Schniepp} Schniepp, H. C. {\it et. al.} {\it J. Phys. Chem. B} {\bf 2006},
{\it 110}, 8535.
\bibitem{Jiang} Jiang, D.; Sumpter, B. G.; Dai, S.; {\it J. Phys Chem. B} {\bf 2006},
{\it 110}, 23628.
\bibitem{Agapito} Agapito, L. A.; Cheng, H-P. {\it J. Phys. Chem. C} {\bf 2007},
{\it 111}, 14266.
\bibitem{Grimme} Grimme, S.; Muck-Lichtenfeld, C.; Antony, J. {\it J. Phys. Chem. C}
{\bf 2007}, {\it 111}, 11199.
\bibitem{Ozyilmaz} Ozyilmaz, B. {\it et. al.} {\it Phys. Rev. Lett.} {\bf 2007}, {\it 99}, 166804.
\bibitem{Berger1} Berger, C. {\it et. al.} {\it Science} {\bf 2006}, {\it 312}, 1191.
\bibitem{Berger2} Berger, C. {\it et al.} {\it J. Phys. Chem. B} {\bf 2004}, {\it 108}, 19912.
\bibitem{Fujita} Fujita, M.; Wakabayashi, K.; Nakada, K.; Kusakabe, K. {\it J. Phys. Soc. Jpn.} 
{\bf 1996}, {\it 65}, 1920.
\bibitem{Nakada} Nakada, K.; Fujita, M.; Dresselhaus, G.; Dresselhaus, M. S.
{\it Phys. Rev. B} {\bf 1996}, {\it 54}, 17954.
\bibitem{Wakabayashi} Wakabayashi, K.; Fujita, M.; Ajiki, H.; Sigrist, M.
{\it Phys. Rev. B} {\bf 1999}, {\it 59}, 8271.
\bibitem{Ezawa} Ezawa, M. {\it Phys. Rev. B} {\bf 2006}, {\it 73}, 045432.
\bibitem{Brey} Brey, L.; Fertig, H. A. {\it Phys. Rev. B} {\bf 2006}, {\it 73}, 235411.
\bibitem{Sasaki} Sasaki, K. I.; Murakami, S.; Saito, R. {\it J. Phys. Soc. Jpn.} {\bf 2006}, 
{\it 75}, 074713.
\bibitem{Abanin} Abanin, D.A.; Lee, P. A.; Levitov, L. S. {\it Phys. Rev. Lett.} {\bf 2006}, 
{\it 96}, 176803.
\bibitem{Neto} Castro Neto, A. H.; Guinea, F.; Peres N. M. R. {\it Phys. Rev. B} {\bf 2006}, 
{\it 73}, 205408.
\bibitem{Sudipta} Dutta, S.; Lakshmi, S.; Pati, S. K. {\it cond-mat/0706.2528v1} {\bf 2007}.
\bibitem{McCann} McCann, E. {\it et. al.} {\it Phys Rev. Lett.} {\bf 2006}, {\it 97}, 146805.
\bibitem{Yang} Yang, L.; Cohen, M. L.; Louie, S. G. {\it Nano Letters} {\bf 2007}, {\it 7}, 3112.
\bibitem{Prezzi} Prezzi, D.; Varsano, D.; Ruini, A.; Marini, A.; Molinari, E.
{\it cond-mat/07060916} {\bf 2007}.
\bibitem{Lin} Hikihara, T.; Hu, X.; Lin, H-H.; Mou, C-Y. {\it Phys. Rev. B} {\bf 2003}, 
{\it 68}, 035432.
\bibitem{Cohen} Son, Y. W.; Cohen, M. L.; Louie, S. G. {\it Nature (London)} {\bf 2006}, 
{\it 444}, 347.; Son, Y. W.; Cohen, M. L.; Louie, S. G. {\it Phys. Rev. Lett.} {\bf 2006}, 
{\it 97}, 216803.
\bibitem{Okada} Okada, S.; Oshiyama, A. {\it Phys. Rev. Lett.} {\bf 2001}, {\it 87}, 146803.
\bibitem{Groot} de Groot, R. A.; Mueller, F. M.; van Engen, P. G.; Buschow, K. H. J.
{\it Phys. Rev. Lett.} {\bf 1983}, {\it 50}, 2024.
\bibitem{Park} Park, J-H. {\it et. al.} {\it Nature} {\bf 1998}, {\it 392}, 794.
\bibitem{Rudberg} Rudberg, E.; Salek, P.; Luo, Y. {\it Nano Letters} {\bf 2007}, {\it 7}, 2211.
\bibitem{Soler} Soler, J. M. {\it et. al.} {\it J. Phys. Condens. Matter} {\bf 2002}, {\it 14}, 2745.
\bibitem{Perdew} Burke, K.; Perdew, J. P.; Ernzerhof, M. {\it Int. J. Quantum Chem.} {\bf 1997}, 
{\it 61}, 287.
\bibitem{Lieb} Lieb, E. H. {\it Phys. Rev. Lett.} {\bf 1989}, {\it 62}, 1201.
\bibitem{Kan} Kan, E-J.; Li, Z.; Yang, J.; Hou, J. G. {\it cond-mat/0708.1213v1} {\bf 2007}.
\bibitem{HRK} Shastry, B. S.; Krishnamurthy, H. R.; Anderson, P. W. {\it Phys. Rev. B}
{\bf 1990}, {\it 41}, 2375.


\end{thebibliography}
\end{document}